\documentclass[letterpaper]{article} 
\usepackage{aaai25}  
\usepackage{times}  
\usepackage{helvet}  
\usepackage{courier}  
\usepackage[hyphens]{url}  
\usepackage{graphicx} 
\usepackage{natbib}  
\usepackage{caption} 
\usepackage{algorithm}
\usepackage{algorithmic}

\usepackage{newfloat}
\usepackage{listings}
\DeclareCaptionStyle{ruled}{labelfont=normalfont,labelsep=colon,strut=off} 
\floatstyle{ruled}
\newfloat{listing}{tb}{lst}{}
\floatname{listing}{Listing}
\usepackage{microtype}
\usepackage{url}
\usepackage{booktabs}

\usepackage{graphicx}
\usepackage{vcell}
\usepackage{multirow}
\usepackage{xcolor}
\usepackage{amsmath}

\title{ResearchCodeAgent: An LLM Multi-Agent System\\ for Automated Codification of Research Methodologies}
\author {
    Shubham Gandhi\textsuperscript{\rm 1},
    Dhruv Shah\textsuperscript{\rm 1},
    Manasi Patwardhan\textsuperscript{\rm 1},
    Lovekesh Vig\textsuperscript{\rm 1},
    Gautam Shroff\textsuperscript{\rm 1}
}
\affiliations {
    \textsuperscript{\rm 1}TCS Research\\
    \{gandhi.shubham, dhruv.bshah2, manasi.patwardhan, lovekesh.vig, gautam.shroff\}@tcs.com
    
}

\usepackage{bibentry}

\lstdefinestyle{plainText}{
    basicstyle=\ttfamily\small,
    backgroundcolor=\color{lightgray},
    frame=single,
    breaklines=true,
    numbers=left,
    numberstyle=\tiny\color{gray},
    stepnumber=1,
    numbersep=5pt,
}

\usepackage{fancyhdr}

\begin{document}

\maketitle

\pagestyle{fancy}
\fancyhf{}
\fancyhead[L]{Presented at the AI4Research workshop @ AAAI 2025}
\fancyhead[R]{Gandhi et al.}
\renewcommand{\headrulewidth}{0.4pt}

\setlength{\headheight}{15pt}
\setlength{\topmargin}{-15pt}
\addtolength{\textheight}{-10pt}
\addtolength{\headsep}{10pt}

\thispagestyle{fancy}

\begin{abstract}
In this paper we introduce ResearchCodeAgent, a novel multi-agent system leveraging large language models (LLMs) agents to automate the codification of research methodologies described in machine learning literature. The system tries to bridge the gap between high-level research concepts and their practical implementation, allowing researchers auto-generating code of existing research papers for benchmarking or building on top-of existing methods specified in the literature with availability of partial or complete starter code. ResearchCodeAgent employs a flexible agent architecture with a comprehensive action suite, enabling context-aware interactions with the research environment. The system incorporates a dynamic planning mechanism, utilizing both short and long-term memory to adapt its approach iteratively. We evaluate ResearchCodeAgent on three distinct machine learning tasks with distinct task complexity and representing different parts of the ML pipeline: data augmentation, optimization, and data batching.  Our results demonstrate the system's effectiveness and generalizability, with 46.9\% of generated code being high-quality and error-free, and 25\% showing performance improvements over baseline implementations. Empirical analysis shows an average reduction of 57.9\% in coding time compared to manual implementation. We observe higher gains for more complex tasks. ResearchCodeAgent represents a significant step towards automating the research implementation process, potentially accelerating the pace of machine learning research. 

\end{abstract}

\section{Introduction}

Research is fundamental to scientific and technological advancement. However, researchers often spend significant time on implementing and codifying methodologies and experiments, reducing the time available for ideation, exploration and designing experiments. This issue is particularly pronounced in the field of Machine Learning (ML), which is characterized by its rapid pace of development and innovation. Given the dynamic nature of ML, it is of prime importance for researchers to receive as much assistance as possible to stay abreast of the latest advancements and maintain a competitive edge. Effective support systems can free researchers from repetitive and time-consuming coding tasks, allowing them to focus on conceptual aspects of their work.

Our work bridges the gap between how a research is conceptualized and how the actual code is implemented. 
Typically, a researcher outlines the methodology, algorithms, and experimental designs at a high-level, abstract manner. Translating these abstractions into working code involves several intricate steps, often requiring deep technical knowledge and significant time investment. While substantial work has been conducted on automation of research problem formulation, \cite{baek2024researchagentiterativeresearchidea} ideation and planning \citep{nigam2024accelerontoolaccelerateresearch,Yang2023LargeLM,wang2024scimonscientificinspirationmachines}, there has been relatively little focus on auto-codifications of these high-level ideas and  experiments. 
The responsibility of implementing ideas suggested by research assistants still largely falls on the researchers themselves. Addressing this gap, an advanced support system such as ResearchCodeAgent tailored to the coding phase can significantly streamline the research process. 

Large Language Models (LLMs) have shown significant improvements in code-related domains, evolving from handling function-level tasks \citep{chen2021evaluating, athiwaratkun2023multilingual} to managing repository-level codebases \citep{bairi2023codeplanrepositorylevelcodingusing,Huang2023MLAgentBenchEL}. Some of the works relies on appropriately defined machine learning (ML) engineering tasks with the availability of code templates \cite{Chan2024MLEbenchEM,Huang2023MLAgentBenchEL}. 
Another parallel works \cite{Li2024MLRCopilotAM} attempts idea, experiment and code generation as sequential tasks. They start with pre-defined machine learning tasks and uses available datasets for these tasks to generate hypothesis followed by generation of experimental design and code. Existing benchmarks such as Scicode \cite{Tian2024SciCodeAR} or BLADE \cite{Gu2024BLADEBL} deal with function level codes for well curated scientific or data-driven problems. SciAgentBench \cite{Chen2024ScienceAgentBenchTR} deal with repository level code but with well-curated task descriptions for the researchers. Implementing research ideas directly from methodological descriptions in research papers is a fundamental yet time-intensive task for researchers. Despite the availability of partial or complete starter code repositories, every researcher must bridge the gap between high-level descriptions and executable implementations to benchmark approaches or build upon existing methods. To the best of our knowledge, no prior work has explicitly addressed the challenge of converting research methodology descriptions into executable code, particularly leveraging repository-level starter code. Every researcher or scientist has to address  this kind of settings for benchmarking her approach on existing approaches or building on top-of existing methods specified in the literature with availability of partial or complete starter code. 

Addressing this realistic scenario, this work aims to facilitate the automated codification of research methodologies and algorithms described in papers. Our goal is to refine the repository level code given inputs such as high-level methodology descriptions, dataset details, and starter code. The importance of starter code cannot be understated, as developing research ideas often involves building on top of previous work.  Focusing on the deep learning domain, we target tasks like data augmentation, optimization, and data batching, restricting our generated code to Python, the predominant language for AI research.


As opposed to well-curated task descriptions, research papers often present methodologies at an abstract level, particularly for complex tasks, which complicates direct codification. To address this, a researcher has to follow a multi-step process including (i) Methodology Decomposition: Break down abstract high-level descriptions into fine-grained sub-tasks, (ii) Starter Code Integration: Identify and refine relevant segments of the provided starter code to implement the sub-tasks and (iii) Iterative Refinement: Execute and validate each sub-task and the overarching methodology, refining the plan as necessary.
Unlike rigid, predefined workflows, this iterative approach mirrors the planning and experimentation process employed by researchers, allowing for adaptive problem-solving based on complexity.


Large Language Model (LLM) code agents are uniquely positioned for such tasks, as they combine environmental understanding with flexible code generation capabilities. Our LLM-based agentic architecture operates in two main phases (i) Context Understanding: Analyzing the environment, including the methodology description, input data, and starter code and (ii) Code Generation and Execution: Planning and generating adaptive, context-aware code to address specific sub-tasks iteratively. This architecture contrasts with single-call LLM baselines by enabling dynamic planning and execution, leading to more accurate and efficient implementations.

We evaluate our architecture on three distinct deep learning tasks described in research papers. Our approach outperforms single-call baselines, producing error-free code more efficiently and saving researchers significant time in editing starter code. Notably, our architecture demonstrates superior performance for complex tasks, highlighting its potential for broader application in automating research implementation workflows.

Through empirical and qualitative analysis, we validate the following claims:
\begin{itemize}
\item  ResearchCodeAgent architecture generalizes effectively across three distinct ML research tasks, each representing different parts of the traditional machine learning pipeline: data augmentation, optimization, and data batching.
\item The system demonstrates a high success rate in code generation, with $46.88\%$ of the generated code being near-perfect and directly usable, $18.75\%$ requiring minor modifications, and $34.38\%$ needing substantial revisions.
\item  Empirical analysis shows that the ResearchCodeAgent yeilds an average reduction of $57.86\%$ in coding efforts as compared to manual implementation. The efficiency gains are observed more for more complex tasks.
\item Our error analysis indicates that the majority of issues encountered by ResearchCodeAgent are related to context understanding and complex logical implementations. However, these errors decrease as the system is iteratively refined, showing an improvement rate of $46.15\%$ over successive trials.
\end{itemize}
\section{ResearchCodeAgent} \label{sec:budgetmlagent}

\begin{figure*}[!ht]
    \centering
    \includegraphics[width=0.88\linewidth]{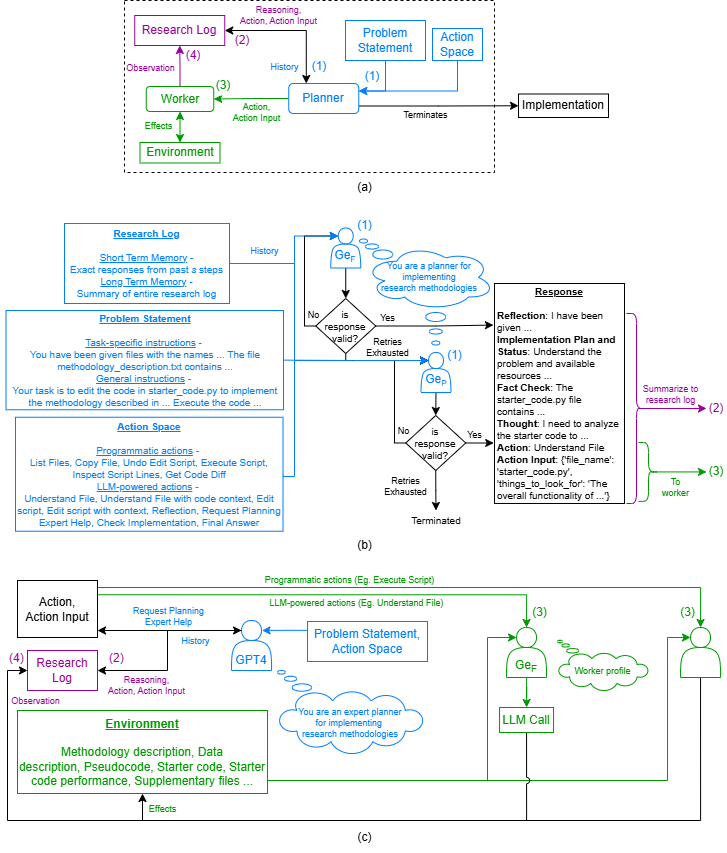}
    \caption{(a) The ResearchCodeAgent system system - \textcolor[HTML]{007FFF}{Planning}, \textcolor[HTML]{990099}{Research Logs}, \textcolor[HTML]{009900}{Workers and Environment} (b) Planning mechanism with LLM Cascade and planner profiles; valid response follows programmatic constraints (c) Planning Expert calls, LLM-powered and programmatic workers.}
    \label{fig:researchcodeagent}
\end{figure*}

In this section we describe our proposed ResearchCodeAgent system, shown in Figure \ref{fig:researchcodeagent}. which interacts with an environment using a suite of actions. Moreover, we also describe its planning mechanism along with programmatic constructs added to aid the system.

\subsection{Environment and inputs}

The input files form the environment that the agent iteratively interacts with. Files here include methodology description, data description, pseudocode, starter code and starter code performance. 
In addition to these mandatory inputs, there might be other files that could be a part of the environment. The starter code script could require other supplementary scripts such as \textit{model.py} to work as intended. For complex methodologies that reference and build upon previous work, we include the original code scripts from the cited papers that correspond to specific methodology sub-parts, similar to how researchers access referenced codebases when implementing methodologies.
The agent is guided by a set of instructions in the form of a problem statement. As shown in appendix listing \ref{lst:listing2}, this problem statement contains both task-specific and generic instructions.

\subsection{Action Space}

As illustrated in Table \ref{tab:actions}, our proposed ResearchCodeAgent system has access to a variety of actions to enable interaction with the environment. This includes programmatic actions viz. `List Files',  Copy File', `Undo Edit Script', `Execute Script', `Inspect Script Lines', `Get Code Diff' and `Final Answer'. Additionally, it also has access to actions that involve underlying LLM calls viz. `Understand File', `Understand File with code context', `Edit Script', `Edit Script with context', `Reflection', `Request Planning Expert Help' and `Check Implementation'. Each of these actions is executed by a worker having a distinct profile or persona that makes the internal LLM calls.

\begin{table}[ht]
\centering
\resizebox{\linewidth}{!}{%
\begin{tabular}{@{}p{4cm}p{5cm}p{7cm}@{}}
\toprule
\textbf{Action Name} & \textbf{Action Inputs} & \textbf{Profile} \\ 
\midrule
List Files & directory path & - \\
Copy File & source, destination & - \\
Inspect Script Lines & script name, start line number, end line number & - \\
Execute Script & script name, arguments & - \\
Undo Edit Script & script name & - \\
Get Code Diff & script 1 name, script 2 name & - \\
Final Answer & description & - \\
Request Planning\\ Expert Help & request description & - \\
Understand File & file name, things to look for & You are an expert in understanding files containing both code and natural language. \\
Understand File\\ with Code Context & file name, file start line number, file end line number, script name, script start line number, script end line number, things to look for & You are an expert in understanding files containing both code and natural language given some context. \\
Edit Script & script name, edit instructions, save script name & You are an expert in editing code files. \\
Edit Script with Context & script name, edit instructions, context file name, file start line number, file end line number, save script name & You are an expert in editing code files given some code or text context. \\
Reflection & things to reflect on & You are an expert in reflecting on previous actions when implementing code for a given research methodology. \\
Check Implementation & script name & You are an expert in checking the implementation of a methodology in a piece of edited code given the starter code that was edited to arrive at the edited code. \\ 
\bottomrule
\end{tabular}%
}
\caption{Programmatic and LLM-driven actions available to ResearchCodeAgent}
\label{tab:actions}
\end{table}

The List Files, Copy File and Inspect Script Lines actions allow the agent system some basic file-level interaction with the environment. 
The Understand File action allows the system to understand a file given a particular set of things to look for. 
The Understand File with code context action does this with an additional snippet of relevant code in context.
The Edit Script action is used to edit a script given a set of edit instructions whereas the Undo Edit Script action is used to undo a previous edit.
The Edit Script with Context action allows the planner to pass an excerpt of text from another file as additional context.
These understand and edit instructions together make up an 'incremental implementation' mechanism which is similar to the way research methodologies are codified in the real world, i.e. in a subpart-by-subpart fashion.
The Execute Script action allows the system to execute a given script. The displayed outputs after execution along with errors, if any, are relayed back as observations. Additionally, it also gives the execution trace as feedback, which contains information about the number of times each line of the given script was executed. 
The Get Code Diff action allows the system to compare an edited code with either the starter code or a previous edited version of the code.

The Reflection action allows the agent system to take a step to gather thoughts and reflect on past actions and observations to make changes to the running plan if required. 
If the planner identifies that it is stuck at a particular step and is unable to proceed, it can make use of the Request Planning Expert Help action which would invoke a much more powerful LLM to plan for that step.
The Check Implementation action allows the system to check the implementation of the methodology in an edited script in a subpart-by-subpart fashion. If a particular subpart is implemented, the worker identifies the snippet of code corresponding to it. If it is not implemented, the worker identifies the snippet of code that needs to be modified to account for subpart implementation and also proposes edits. Lastly, the Final Answer action can be used by the planner to end the run once it is convinced that it is appropriate to do so.

\subsection{Planning mechanism}

At each step, the planner is given the entire responses for the past few steps in context which makes up the short-term memory of the planner. Additionally, it is also given a summarized version of the entire history of the research log up until that step, which makes up the long-term memory.
Similar to the workers, the planner also has a distinct persona. At each step, the planner needs to adhere to a particular response format -  Reflection, Research Plan and Status, Fact Check, Thought, Action and Action Input. This structured response makes the plan implicitly 'running' by nature, i.e., the agent does not start off with a fixed plan given its limited context in the beginning, but instead adapts at each step based on observations.
At each step, a summarized version of the reasoning, action and observation is appended to the research log to serve as part of the long-term memory. Moreover, if the observation at a particular step is too long for short-term memory, it is further summarized using LLM calls.
The planner is given a fixed number of retries at each step for generating a valid response. In case the response by the base planner is not valid even after exhausting the max retries for that level, the system invokes a more powerful, but more costly, LLM for planning for that step. We refer to this mechanism as an LLM cascade.

\subsection{Programmatic Constructs}

Preliminary runs revealed a few mistakes that the system made repeatedly, in both the planner and worker-related phases. 
To aid the system in overcoming these shortcomings, we introduce some programmatic constructs. LLM Agents, in general, tend to loop over actions (add references here) and select similar actions repeatedly until they terminate. To avoid such looping in ResearchCodeAgent, we introduce a pooling mechanism wherein we divide the action space into three pools. 
Pool A contains understanding and planning-related actions such as 'List Files', 'Understand File', 'Understand File with code context', 'Inspect Script Lines', 'Get Code Diff'. Pool B contains code-related actions such as 'Copy File', 'Undo Edit Script', 'Execute Script', 'Edit Script (AI)', 'Edit Script (AI) with context'. The rest of the actions make up Pool C, a more general set of actions.
We prevent the planner from selecting an action from the same pool A or B consecutively for k number of steps. Since there might be a need for more consecutive similar actions towards the beginning, this maximum limit k exponentially decays over steps. Since the actions in Pool C are more general in nature, we do not impose such constraints on the planner calling them.
Additional programmatic constructs include the following - avoiding consecutive duplicate actions, avoiding recursive non-terminating responses and avoiding zero-diff when edit script action is called.

\section{Experimentation}

In this section, we describe the experimental setup for applying ResearchCodeAgent to diverse research tasks with varying complexities in methodology and code.

\subsection{Case studies} \label{sec:datapoints}
We demonstrate the generalizability of the ResearchCodeAgent system by applying it to a diverse set of research tasks spanning different stages of a traditional machine learning pipeline. These tasks vary significantly in complexity, both in terms of methodology and codebase. The works mentioned in following subsections arein increasing order of complexity of code and methodology, viz. OGSCL \citep{Paliwal2023OntologyGS} deals with data batching, YONA \citep{hu2024needhalfboostingdata} deals with data augmentation and FLAG \citep{kong2022robustoptimizationdataaugmentation} deals with optimization. We measure the difficulty of the code in terms of the number of edits that ideally need to be made on the starter code to implement the methodology described in the research paper.
Following is a description of each of the data points we consider.


\subsubsection{OGSCL}

OGSCL is a methodology designed for fine-tuning the CLIP \citep{pmlr-v139-radford21a} model on the DeepFashion dataset. The key idea behind OGSCL is to ensure that each batch during training consists of samples with only one attribute type. The DeepFashion dataset, which is used in this methodology, contains images of fashion apparel products along with their labels. Each product image is annotated with a product category (PC) and attribute values (AV) for up to five different attribute types (AT): fabric, style, shape, texture, and part. A typical label for a product is formatted as [PC, (`fabric', AV1), (`style', AV2), (`shape', AV3), (`texture', AV4), (`part', AV5)]. Not every image is labeled with all five attribute types, but the dataset encompasses a total of 983 attribute values.
For the purposes of OGSCL, the DeepFashion dataset \citep{7780493} is transformed into an image captioning dataset. This transformation involves creating a text caption for each (AT, AV) pair using the format: "The $<$AT$>$ of this $<$PC$>$ is $<$AV$>$." As a result, each image in the dataset is repeated as many times as the number of (AT, AV) pairs it contains. In the transformed dataset, each data point consists of an image and a corresponding caption that describes a specific attribute value for one attribute type of the product. The modified DeepFashion dataset used in our experiments contains 500 samples, with 100 samples for each attribute type. The CLIP-Base model was used for fine-tuning, with training limited to a single epoch due to computational constraints. The starter code for implementing this methodology was developed entirely from scratch.

\subsubsection{YONA}
\citet{hu2024needhalfboostingdata} proposed You Only Need hAlf (YONA), an algorithm for data augmentation on images aimed at improving model robustness by removing redundant information. The technique involves slicing an image randomly into halves, converting one side to noise, applying augmentations (e.g., flips) to the other side, and combining the halves for training.
To evaluate the capability of our ResearchCodeAgent system in replicating the methodology described in the YONA paper, we utilized their publicly available implementation\footnote{\url{https://github.com/yuncheng97/YONA}}. This repository provides code for training various ResNet models \citep{he2015deepresiduallearningimage} on the CIFAR-10 and CIFAR-100 datasets. For our baseline, we selected ResNet-18 and the CIFAR-10 dataset, modifying the original code to remove all references to the YONA methodology. To accommodate computational constraints and focus on implementing the augmentation method, we reduced the number of training epochs to one. The performance of this starter code was included as part of the input files in the environment.
The YONA methodology was directly incorporated from the paper in verbatim, while the dataset description was refined based on information from the CIFAR dataset website \footnote{\url{https://www.cs.toronto.edu/~kriz/cifar.html}}. Since the paper did not provide explicit pseudocode, we constructed a pseudocode representation for the YONA algorithm to aid in its implementation.

\subsubsection{FLAG}

\citet{kong2022robustoptimizationdataaugmentation} introduce the Free Large-scale Adversarial Augmentation on Graphs (FLAG) algorithm, a sophisticated technique designed to enhance the performance of Graph Neural Networks (GNNs). FLAG leverages adversarial perturbation, projected gradient descent, multi-scale augmentation, and free training to iteratively augment node features with gradient-based adversarial perturbations during the training process. This iterative augmentation helps the model achieve invariance to small input data fluctuations, thereby improving generalization to out-of-distribution samples and boosting performance during testing.

To evaluate the capability of our ResearchCodeAgent system in replicating the methodology described in the FLAG paper, we utilized the publicly available implementation of FLAG\footnote{\url{https://github.com/devnkong/FLAG}}. Our experiments focus on the ogbn-arxiv dataset, employing Graph Convolutional Network (GCN) \citep{kipf2017semisupervisedclassificationgraphconvolutional} and GraphSAGE \citep{hamilton2018inductiverepresentationlearninglarge} models. The FLAG repository provides baseline code for these models using vanilla training, which we adopted as starter code, after reducing the number of epochs to 50, owing to computational constraints. We also include the performance of this starter code as part of the input files in the environment.
We incorporated the verbatim methodology description from the FLAG paper, the data description from the ogbn-arxiv webpage\footnote{\url{https://ogb.stanford.edu/docs/nodeprop/\#ogbn-arxiv}}, and a refined version of the pseudocode presented in the paper. Additionally, the original FLAG paper includes citations for various methodological subparts; we included the complete scripts from these citations as part of our environment to provide supplementary information to the system.

\subsection{Baselines}

We evaluate our proposed ResearchCodeAgent system by comparing it with two baseline approaches:

\subsubsection{Prescribed path}

In this baseline, we try to constrain the system to follow a predetermined sequence of actions. This approach involves adding specific instructions to the problem statement that outline a step-by-step process for implementing the methodology. The prescribed plan includes steps such as listing and understanding input files, creating a skeleton with function definitions for methodology subparts, implementing each function sequentially, checking the implementation of each subpart, and using specific actions in a prescribed order. This baseline allows us to assess the value of the adaptive planning mechanism in ResearchCodeAgent by comparing it to a more rigid, predefined approach. By comparing ResearchCodeAgent against this baseline, we aim to demonstrate the benefits of flexible, adaptive planning over a fixed action sequence, the importance of context-aware decision-making in complex coding tasks, and the impact of allowing the agent to dynamically adjust its approach based on intermediate results and observations.

\subsubsection{Single LLM Call}

For this baseline, we utilize a single LLM call to generate the entire implementation in one pass. Given the same inputs as ResearchCodeAgent (methodology description, data description, pseudocode, starter code and other files), the model is prompted to produce a complete implementation of the research methodology. This approach tests whether the iterative, multi-agent system of ResearchCodeAgent offers advantages over a simpler, one-shot generation method.

\subsection{Metrics}

To evaluate the effectiveness of ResearchCodeAgent, we employ several metrics that assess both the quality of the generated code and the efficiency of the system. 

\subsubsection{Code Quality Metrics}

\begin{enumerate}

    \item \textbf{Category-wise Distribution}: We divide the generated code into four categories based on its functionality and performance:
    \begin{itemize}
        \item A. Error-free with performance improvement: Code that runs without errors and demonstrates improved performance compared to the baseline starter code.
        \item B. Error-free without performance improvement: Code that runs without errors but does not show significant performance gains.
        \item C. Erroneous code: Code that contains errors and fails to execute properly.
        \item D. Terminated without generating code: Cases where the system fails to produce any code.
    \end{itemize}
    Analyzing category proportions helps evaluate the approach's effectiveness in generating functional, performance-enhancing research implementations.
    

    \item \textbf{Code quality}: We further classify the code quality into three bins based on manual scores assigned by two expert reviewers:

    \begin{itemize}
        \item S1 (8-10): High-quality code requiring minor or no repairs. 
        \item S2 (4-7): Code requiring major repairs but demonstrating a partial understanding of the methodology. 
        \item S3 (1-3): Low-quality code with serious flaws or misinterpretations of the methodology. 
    \end{itemize}

    This score-wise distribution provides a more nuanced view of the code quality, allowing us to assess not just the success rate but also the degree of correctness and usability of the generated code.
    
\end{enumerate}

\subsubsection{Efficiency Metrics}

\begin{enumerate}
    \item \textbf{Average Lines Edited}: This metric measures the average number of lines modified by ResearchCodeAgent during the implementation process.

    \item \textbf{Average Lines Repaired}: This metric indicates the average number of lines that required manual correction after the system's implementation.

    \item \textbf{Average Time Saving}: To quantify the reduction in coding effort, we calculate the percentage of time saved compared to manual implementation using the formula:
    $$\text{Time Saving (\%)} = (1 - \frac{\text{Lines Repaired}}{\text{Lines Edited}}) \times 100$$

\end{enumerate}

\subsection{Models and hyperparameters}

ResearchCodeAgent employs a cascade of Large Language Models (LLMs) for its planning and execution mechanisms. The base planner utilizes Gemini 1.5 Flash\footnote{\url{https://ai.google.dev/gemini-api/docs/models/gemini#gemini-1.5-flash}}, with Gemini 1.5 Pro\footnote{\url{https://ai.google.dev/gemini-api/docs/models/gemini#gemini-1.5-pro}} as the intermediate planner in the cascade. For challenging planning steps, the system escalates to GPT-4\footnote{\textit{gpt-4-0125-preview} - \url{https://platform.openai.com/docs/models}} as the planning expert. All intelligent workers that require LLM capabilities are powered by Gemini 1.5 Flash.
To balance exploration and consistency, we use different temperature settings for various components of the system. Planning-related calls employ a higher temperature of 0.8 to encourage creative problem-solving, while worker calls use a lower temperature of 0.2 for more deterministic outputs.
The planning mechanism integrates short- and long-term memory. Short-term memory uses a context window containing responses from the last three steps to maintain recent context. The planner permits up to three planning expert calls per run, reserving the strongest model for critical decisions. For LLM calls, retries are limited to 8 for Gemini 1.5 Flash, 4 for Gemini 1.5 Pro, and 1 for GPT-4..
To prevent action looping, we employ a dynamic pooling mechanism. Initially, the system allows a maximum of 15 consecutive actions from the same pool, with this limit decaying exponentially at a rate of 0.01 as the implementation progresses. This approach ensures a diverse action selection while allowing for necessary repetition in the early stages of code generation.
These hyperparameters and constraints can be fine-tuned on a case-by-case basis to optimize performance for different research tasks and methodologies, demonstrating the flexibility of the ResearchCodeAgent architecture. The Appendix shows the problem statements, dataset descriptions, methodology descriptions and pseudocodes for the cases we consider.

\begin{table}[!ht]
\centering
\resizebox{\columnwidth}{!}{%
\begin{tabular}{@{}c l cccc@{}}
\toprule
\textbf{Experiment} & \textbf{Datapoint} & \textbf{A (\%)} & \textbf{B (\%)} & \textbf{C (\%)} & \textbf{D (\%)} \\ 
\midrule
\multirow{5}{*}{\textbf{\begin{tabular}[c]{@{}c@{}}Single LLM\\ Call\end{tabular}}} 
    & FLAG - GCN & 0 & 0 & 100 & 0 \\ 
    & FLAG - GraphSAGE & 0 & 0 & 100 & 0 \\ 
    & YONA & 0 & 25 & 75 & 0 \\ 
    & OGSCL & 50 & 0 & 50 & 0 \\ 
\cmidrule{2-6}
    & \textbf{Average} & \textbf{12.5} & \textbf{6.25} & \textbf{81.25} & \textbf{0} \\
\midrule
\multirow{5}{*}{\textbf{\begin{tabular}[c]{@{}c@{}}Prescribed Plan\end{tabular}}} 
    & FLAG - GCN & 0 & 50 & 50 & 0 \\ 
    & FLAG - GraphSAGE & 0 & 25 & 75 & 0 \\ 
    & YONA & 0 & 0 & 100 & 0 \\ 
    & OGSCL & 0 & 50 & 25 & 25 \\ 
\cmidrule{2-6}
    & \textbf{Average} & \textbf{0} & \textbf{31.25} & \textbf{62.5} & \textbf{6.25} \\
\midrule
\multirow{5}{*}{\textbf{ResearchCodeAgent}} 
    & FLAG - GCN & \textbf{12.5} & \textbf{37.5} & \textbf{50} & \textbf{0} \\ 
    & FLAG - GraphSAGE & \textbf{37.5} & \textbf{12.5} & \textbf{37.5} & \textbf{12.5} \\ 
    & YONA & 12.5 & 37.5 & 37.5 & 12.5 \\ 
    & OGSCL & 37.5 & 0 & 50 & 12.5 \\ 
\cmidrule{2-6}
    & \textbf{Average} & \textbf{25} & \textbf{21.88} & \textbf{43.75} & \textbf{9.38} \\ 
\bottomrule
\end{tabular}%
}
\caption{Error Categories Analysis. A: Error-free with performance improvement, B: Error-free w/o performance improvement, C: Erroneous code, D: Terminated without generating code}
\label{tab:error-categories}
\end{table}

\begin{table}[!ht]
\centering
\resizebox{\columnwidth}{!}{%
\begin{tabular}{@{}c l ccc@{}}
\toprule
\textbf{Experiment} & \textbf{Datapoint} & \textbf{S1 (\%)} & \textbf{S2 (\%)} & \textbf{S3 (\%)} \\ 
\midrule
\multirow{5}{*}{\textbf{\begin{tabular}[c]{@{}c@{}}Single LLM\\ Call\end{tabular}}} 
    & FLAG - GCN & 0 & 0 & 100 \\ 
    & FLAG - GraphSAGE & 0 & 25 & 75 \\ 
    & YONA & 50 & 50 & 0 \\ 
    & OGSCL & 100 & 0 & 0 \\ 
\cmidrule{2-5}
    & \textbf{Average} & \textbf{37.5} & \textbf{18.75} & \textbf{43.75} \\
\midrule
\multirow{5}{*}{\textbf{\begin{tabular}[c]{@{}c@{}}Prescribed Plan\end{tabular}}} 
    & FLAG - GCN & 0 & 0 & 100 \\ 
    & FLAG - GraphSAGE & 0 & 25 & 75 \\ 
    & YONA & 0 & 25 & 75 \\ 
    & OGSCL & 0 & 50 & 50 \\ 
\cmidrule{2-5}
    & \textbf{Average} & \textbf{0} & \textbf{25} & \textbf{75} \\
\midrule
\multirow{5}{*}{\textbf{ResearchCodeAgent}} 
    & FLAG - GCN & \textbf{37.5} & \textbf{25} & \textbf{37.5} \\ 
    & FLAG - GraphSAGE & \textbf{37.5} & \textbf{12.5} & \textbf{50} \\ 
    & YONA & 37.5 & 25 & 37.5 \\ 
    & OGSCL & 75 & 12.5 & 12.5 \\ 
\cmidrule{2-5}
    & \textbf{Average} & \textbf{46.88} & \textbf{18.75} & \textbf{34.38} \\ 
\bottomrule
\end{tabular}%
}
\caption{Code Quality Analysis. S1 (8-10): Good code requiring minor repairs, S2 (4-7): Code requiring major repairs, S3 (1-3): Bad code with serious flaws}
\label{tab:code-quality}
\end{table}

\begin{table}[!ht]
\centering
\resizebox{\columnwidth}{!}{%
\begin{tabular}{@{}c l c c c@{}}
\toprule
\textbf{Experiment} & \textbf{Datapoint} & \textbf{\begin{tabular}[c]{@{}c@{}}Avg. \#Lines\\ Edited\end{tabular}} & \textbf{\begin{tabular}[c]{@{}c@{}}Avg. \#Lines\\ Repaired\end{tabular}} & \textbf{\begin{tabular}[c]{@{}c@{}}Avg. Time\\ Saving (\%)\end{tabular}} \\ 
\midrule
\multirow{5}{*}{\textbf{\begin{tabular}[c]{@{}c@{}}Single LLM\\ Call\end{tabular}}} 
    & FLAG - GCN & 47.25 & 25.00 & 46.11 \\ 
    & FLAG - GraphSAGE & 53.00 & 25.75 & 51.82 \\ 
    & YONA & 67.75 & 45.00 & 57.39 \\ 
    & OGSCL & 17.50 & 0.75 & 95.48 \\ 
\cmidrule{2-5}
    & \textbf{Average} & \textbf{46.38} & \textbf{24.63} & \textbf{62.20} \\
\midrule
\multirow{5}{*}{\textbf{\begin{tabular}[c]{@{}c@{}}Prescribed Plan\end{tabular}}} 
    & FLAG - GCN & 48.25 & 26.50 & 44.75 \\ 
    & FLAG - GraphSAGE & 77.25 & 33.50 & 44.38 \\ 
    & YONA & 40.00 & 35.75 & 10.52 \\ 
    & OGSCL & 28.25 & 17.00 & 28.06 \\ 
\cmidrule{2-5}
    & \textbf{Average} & \textbf{48.94} & \textbf{28.69} & \textbf{31.93} \\
\midrule
\multirow{5}{*}{\textbf{ResearchCodeAgent}} 
    & FLAG - GCN & \textbf{45.63} & \textbf{9.38} & \textbf{77.82} \\ 
    & FLAG - GraphSAGE & \textbf{55.00} & \textbf{11.75} & \textbf{63.02} \\ 
    & YONA & 31.00 & 23.88 & 31.93 \\ 
    & OGSCL & 25.50 & 13.25 & 58.66 \\ 
\cmidrule{2-5}
    & \textbf{Average} & \textbf{39.78} & \textbf{14.57} & \textbf{57.86} \\ 
\bottomrule
\end{tabular}%
}
\caption{Performance Metrics Analysis}
\label{tab:performance-metrics}
\end{table}

\section{Results and Discussion}

Our evaluation of ResearchCodeAgent aims to answer several key research questions about its effectiveness, efficiency, and generalizability across machine learning tasks varying in terms of methodology and complexity. We analyze the results for three distinct ML tasks: data batching (OGSCL), data augmentation (YONA) and optimization (FLAG). We perform 8 runs for each task, and also consider two variations for FLAG with GCN and GraphSAGE.

\subsection{RQ1: How is the correctness and quality of the code generated by ResearchCodeAgent?}

Table \ref{tab:error-categories} and Table \ref{tab:code-quality} illustrate ResearchCodeAgent's significant advantages over Single LLM Call and Prescribed Plan methods in both error minimization and code quality respectively. When examining the error categories, ResearchCodeAgent consistently reduces the rate of errors. On average, it produces fewer erroneous outputs (C) at 43.75\%, compared to 62.5\% for the Prescribed Plan and 81.25\% for the Single LLM Call. Moreover, ResearchCodeAgent also consistently produces error-free code with improvement in performance as compared to other approaches. 
From a code quality perspective, ResearchCodeAgent achieves the highest proportion of S1 outputs, at 46.88\%, outperforming  Prescribed Plan (0\%) and Single LLM Call (37.5\%). It also generates less low-quality S3 code, at 34.38\%, significantly better than Prescribed Plan (75\%) and Single LLM Call (43.75\%). These findings suggest that ResearchCodeAgent produces outputs that are not only less erroneous but also closer to implementation-ready code in terms of quality.


\subsection{RQ2: How does ResearchCodeAgent compare to the Prescribed Plan approach?}

The results in Tables \ref{tab:error-categories}, \ref{tab:code-quality} and \ref{tab:performance-metrics} demonstrate ResearchCodeAgent’s consistent superiority over the Prescribed Plan across all metrics: error categories, code quality scores, and time savings. ResearchCodeAgent generates significantly more error-free outputs (A + B categories) at 46.88\%, compared to 31.25\% for the Prescribed Plan, and reduces erroneous outputs (C category) to 43.75\%, markedly better than the Prescribed Plan (62.5\%). It also minimizes bad-quality outputs requiring significant repairs (S3 bin) to 34.38\%, a notable improvement over the Prescribed Plan (75\%), while maximizing S1 quality code from 0\% to 46.88\%. In terms of time saving, ResearchCodeAgent achieves an average time saving of 57.86\%, significantly higher than the Prescribed Plan (31.93\%). All of these trends are consistent across tasks of all difficulty levels. 
These results emphasize ResearchCodeAgent’s adaptability and efficiency, showcasing its ability to handle diverse tasks better than the restrictive and less flexible Prescribed Plan.

\subsection{RQ3: How does task complexity affect time saved in code implementation?}

ResearchCodeAgent demonstrates distinct time saving across tasks of varying complexity. Table \ref{tab:performance-metrics} shows that the system achieves an average time saving of 57.86\%, slightly below the Single LLM Call (62.20\%) but significantly higher than the Prescribed Plan (51.93\%). Its advantages are most evident in high-complexity tasks, such as FLAG, where dynamic planning results in substantial time savings: 77.82\% for FLAG-GCN and 63.02\% for FLAG-GraphSAGE. This highlights the strength of ResearchCodeAgent in dynamically adapting to challenging scenarios, enabling more efficient and effective problem-solving than the Prescribed Plan or Single LLM Call approaches.
In contrast, for simpler tasks like YONA and OGSCL, the Single LLM Call approach consistently outperforms, achieving better time savings. This suggests that the overhead introduced by ResearchCodeAgent’s dynamic planning may not be justified for straightforward problems, where static solutions are already optimized for efficiency. For instance, YONA shows 31.93\% time saving with ResearchCodeAgent compared to a higher efficiency by Single LLM Call (57.39\%). This divergence points to the trade-offs inherent in using adaptive systems for tasks where simplicity outweighs adaptability.

These findings emphasize that task complexity is a critical determinant of ResearchCodeAgent’s relative effectiveness. The system’s design excels in scenarios demanding iterative refinement and dynamic adjustments but introduces unnecessary complexity for simpler tasks. Thus, deploying ResearchCodeAgent optimally involves aligning its capabilities with the complexity of the tasks at hand.

\section{Conclusion}

ResearchCodeAgent represents a significant advancement in automating the codification of research methodologies in machine learning. By employing a multi-agent LLM system with dynamic planning, it consistently demonstrates superiority over static approaches like the Prescribed Plan and simpler Single LLM Call strategies, particularly for complex tasks.
Our evaluation highlights several key strengths of ResearchCodeAgent. The system produces high-quality, research-grade code with a substantial proportion (46.9\%) of outputs being directly usable or requiring only minor modifications, outperforming other approaches in terms of error minimization and quality improvement. 
ResearchCodeAgent's dynamic planning mechanism enables it to handle tasks of varying complexity with notable efficiency. It achieves an average time saving of 57.86\%, surpassing the Prescribed Plan's 31.93\% and approaching the efficiency of Single LLM Call for straightforward tasks. This adaptability is especially advantageous for high-complexity tasks, where the system excels in delivering efficient, high-quality solutions. However, for simpler tasks, the overhead introduced by dynamic planning can result in suboptimal performance compared to static solutions.
Overall, ResearchCodeAgent demonstrates considerable potential to accelerate the research process by bridging the gap between conceptual descriptions and practical code implementations.

\bibliography{aaai25}

\newpage
\onecolumn
\appendix

\UseRawInputEncoding
\section{Appendix}

\begin{listing}[!h]%
\caption{FLAG - Prompt for Single LLM Call}%
\label{lst:listing2}%
\begin{lstlisting}
ETHODOLOGY DESCRIPTION: 

{methodology}

STARTER CODE:

{starter_code}

DATASET DESCRIPTION:

{data}

PSEUDOCODE:

{pseudocode}

STARTER CODE PERFORMANCE:

{starter_code_performance}

SUBPART_1_a CODE:

{subpart_1_a}

SUBPART_2_a CODE:

{subpart_2_a}

Edit the starter code for implementing the methodology using the information given above.
\end{lstlisting}
\end{listing}


\begin{listing}[!h]%
\caption{FLAG - Problem statement for Prescribed Path approach}%
\label{lst:listing1}%
\begin{lstlisting}
List the files in the working directory to check which of the above listed files are actually given. This will also help you identify additional files.
Go through these files and understand their contents.
The file methodology_description.txt contains a detailed description of the methodology.
The file pseudocode.txt contains a pseudocode representing the methodology. 
The file dataset_description.txt contains a detailed description of the dataset. Go through the dataset_description.txt file to understand the data and all the features.
The file starter_code.py contains a starter code. You will need to first understand what this code does and also identify where you need to make changes to implement the methodology. 
The file starter_code_performance.txt contains the performance of the starter code.
The working directory may also contain additional files named subpart_[i]_[j].py, which represents the j'th code script related to the i'th subpart of the methodology. Do not try to execute these subpart scripts. Ignore if such files are not present. 
You can summarize the methodology_description.txt, dataset_description.txt, pseudocode.txt, starter_code.py, starter_code_performance.txt and any other additional files in your research logs to keep track of what all you have to do.
Your task is to edit the code in starter_code.py to implement the methodology described in methodology_description.txt and pseudocode.txt for the dataset described in dataset_description.txt.
Do not forget to execute the changes you made to check for performance. The edited code should have performance greater than what is mentioned in starter_code_performance.txt. Save the edited script in a new file named methodology_implementation.py.
Do not make any changes to the methodology. Do not change any arguments or hyperparameters from the starter code.
After listing and understanding files, when you start editing the code, first add a skeleton containing function definitions for methodology subparts. Then implement each of these functions and check the implementation. Use the Understand File with code context and Edit Script with context actions to implement the subparts by giving appropriate context.
Make sure that you make appropriate edits to the code and execute the edited code and not just execute unedited code to submit final answer.
Repair any errors that the code might have. Do not submit code with errors. Execute the code to verify that it does not have errors.
Use the Check Implementation action to check if the python script contains implementation of the methodology before submitting final answer and that the performance of the final code is greater than the performance of starter_code.py mentioned in starter_code_performance.txt. 
Verify that the part of the code related to the methodology implementation is actually being called during execution using the execution trace files obtained in <script_name>_execution_trace.covers which contain the number of times each line is executed and >>>>>> denotes that the line is not executed even once.
\end{lstlisting}
\end{listing}

\begin{listing}[!h]%
\caption{FLAG - Problem statement for ResearchCodeAgent}%
\label{lst:listing2}%
\begin{lstlisting}
You have been given files with the names methodology_description.txt, dataset_description.txt, pseudocode.txt,starter_code.py and starter_code_performance.txt.
List the files in the working directory to check which of the above listed files are actually given. This will also help you identify additional files.
Go through these files and understand their contents.
The file methodology_description.txt contains a detailed description of the methodology.
The file pseudocode.txt contains a pseudocode representing the methodology. 
The file dataset_description.txt contains a detailed description of the dataset. Go through the dataset_description.txt file to understand the data and all the features.
The file starter_code.py contains a starter code. You will need to first understand what this code does and also identify where you need to make changes to implement the methodology. 
The file starter_code_performance.txt contains the performance of the starter code.
The working directory may also contain additional files named subpart_[i]_[j].py, which represents the j'th code script related to the i'th subpart of the methodology. Do not try to execute these subpart scripts. Ignore if such files are not present. 
You can summarize the methodology_description.txt, dataset_description.txt, pseudocode.txt, starter_code.py, starter_code_performance.txt and any other additional files in your research logs to keep track of what all you have to do.
Your task is to edit the code in starter_code.py to implement the methodology described in methodology_description.txt and pseudocode.txt for the dataset described in dataset_description.txt.
Do not forget to execute the changes you made to check for performance. The edited code should have performance greater than what is mentioned in starter_code_performance.txt. Save the edited script in a new file named methodology_implementation.py.
Do not make any changes to the methodology. Do not change any arguments or hyperparameters from the starter code.
Make sure that you make appropriate edits to the code and execute the edited code and not just execute unedited code to submit final answer.
Repair any errors that the code might have. Do not submit code with errors. Execute the code to verify that it does not have errors.
Use the Check Implementation action to check if the python script contains implementation of the methodology before submitting final answer and that the performance of the final code is greater than the performance of starter_code.py mentioned in starter_code_performance.txt. 
Verify that the part of the code related to the methodology implementation is actually being called during execution using the execution trace files obtained in <script_name>_execution_trace.cover which contain the number of times each line is executed and >>>>>> denotes that the line is not executed even once. 
\end{lstlisting}
\end{listing}


\begin{listing}[!h]%
\caption{FLAG - Dataset Description}%
\label{lst:listing2}%
\begin{lstlisting}
Graph: The ogbn-arxiv dataset is a directed graph, representing the citation network between all Computer Science (CS) arXiv papers indexed by MAG. Each node is an arXiv paper and each directed edge indicates that one paper cites another one. Each paper comes with a 128-dimensional feature vector obtained by averaging the embeddings of words in its title and abstract. The embeddings of individual words are computed by running the skip-gram model over the MAG corpus. We also provide the mapping from MAG paper IDs into the raw texts of titles and abstracts here. In addition, all papers are also associated with the year that the corresponding paper was published.

Prediction task: The task is to predict the 40 subject areas of arXiv CS papers, e.g., cs.AI, cs.LG, and cs.OS, which are manually determined (i.e., labeled) by the paper’s authors and arXiv moderators. With the volume of scientific publications doubling every 12 years over the past century, it is practically important to automatically classify each publication’s areas and topics. Formally, the task is to predict the primary categories of the arXiv papers, which is formulated as a 40-class classification problem.

Dataset splitting: We consider a realistic data split based on the publication dates of the papers. The general setting is that the ML models are trained on existing papers and then used to predict the subject areas of newly-published papers, which supports the direct application of them into real-world scenarios, such as helping the arXiv moderators. Specifically, we propose to train on papers published until 2017, validate on those published in 2018, and test on those published since 2019.
\end{lstlisting}
\end{listing}


\begin{listing}[!h]%
\caption{FLAG - Methodology Description}%
\label{lst:listing2}%
\begin{lstlisting}
Following is the description of the FLAG (Free Large-scale Adversarial Augmentation on Graphs) algorithm.
In this work, we investigate how to effectively improve the generalization of GNNs through a feature based augmentation. Graph node features are usually constructed as discrete embeddings, such as binary bag-of-words vectors or categorical variables. As a result, standard hand-crafted augmentations, like flipping and cropping transforms used in computer vision, are not applicable to graphs node features.
By hunting for and stamping out small perturbations that cause the classifier to fail, one may hope that adversarial training could benefit standard accuracy (Goodfellow et al., 2014; Tsipras et al., 2018; Miyato et al., 2018). It is widely observed that when the data distribution is sparse and discrete, the beneficial effect of adversarial perturbations on generalization takes over (Tsipras et al., 2018; Gan et al., 2020). Volpi et al. (2018) viewed adversarial perturbation as a data-dependent regularization, which could intuitively generalize to out-of-distribution samples. Highlighted by Hu et al. (2020), the out-of-distribution phenomenon of data is salient in the graph domain, and also considering the sparsity of labeled node samples in the semi-supervised node classi cation task, we view adversarial perturbation as a strong candidate method for input feature augmentation.
Min-Max Optimization. Adversarial training is the process of crafting adversarial data points, and then injecting them intro training data. This process is often formulated as the following min-max problem:
Eq (3) - The equation represents the minimization over theta of the expected value, where the expectation is taken over pairs of x and y drawn from the distribution D. The quantity being minimized is the maximum value of the loss function L, evaluated at points f sub theta of x plus delta, and y, where the maximum is taken over all delta such that the p-norm of delta is less than or equal to epsilon.
where D is the data distribution, y is the label, || . || p is some p-norm distance metric, epsilon is the perturbation budget, and L is the objective function. Madry et al. (2017) showed that this saddle point optimization problem could be reliably tackled by Stochastic Gradient Descent (SGD) for the outer minimization and Projected Gradient Descent (PGD) for the inner maximization. In practice, the typical approximation of the inner maximization under an l-norm constraint is as follows, 
Eq (4) - Delta at time t plus 1 is equal to the projection onto the set of delta for which the infinity norm of delta is less than or equal to epsilon of the quantity delta sub t plus alpha times the sign of the gradient of the loss function L evaluated at f sub theta of x plus delta sub t, y.
where the perturbation delta is updated iteratively, and performs projection onto the-ball in the l-norm. For maximum robustness, this iterative updating procedure usually loops M times to craft the worst-case noise, which requires M forward and backward passes end-to-end. Afterwards the most vicious noise is applied to the input feature, on which the model weight is optimized. The algorithm above is called PGD.
Multi-scale Augmentation. On visual tasks, Chen et al. (2020) highlighted the importance of using diverse types of data augmentations such as random cropping, color distortion, and Gaussian blur. The authors showed that a single transformation is not su cient to learn good representations. To fully exploit the generalizing ability and enhance the diversity and quality of adversarial perturbations, we propose to craft multi-scale augmentations. To realize this goal, we leverage the techniques below.
Free training. We leverage free adversarial training (Shafahi et al., 2019) to craft adversarial data augmentations. PGD is a powerful yet ine cient way of solving the min-max optimization. It runs M full forward and backward passes to craft a refined perturbation delta 1:M, but the model weights theta only get updated once using the final delta M. This process makes model training M times slower. In contrast, while computing the gradient for the perturbation , free training simultaneously produces the model parameter on the same backward pass. This enables a parameter update to be computed in parallel with a perturbation update at virtually no additional cost. The authors proposed to train on the same minibatch M times in a row to simulate the inner maximization in Eq. (3), while compensating by performing M times fewer epochs of training. The resulting algorithm yields accuracy and robustness competitive with standard adversarial training, but with the same runtime as clean training.
....
\end{lstlisting}
\end{listing}


\begin{listing}[!h]%
\caption{FLAG - Pseudocode}%
\label{lst:listing2}%
\begin{lstlisting}
FLAG Algorithm in Pseudocode/Natural Language with Symbol Explanations:
Input:
Graph G = (V, E) with labeled node set Vl and unlabeled node set Vu.
V represents the set of all nodes in the graph.
E represents the set of edges connecting nodes.
Vl is the subset of nodes with known labels.
Vu is the subset of nodes without known labels.
Learning rate (tau).
Ascent steps M: the number of iterations for updating parameters.
Ascent step size (alpha_l) for labeled nodes and (alpha_u) for unlabeled nodes.
Objective function L: a function to measure the model's performance.
Aggregation function A: a function to combine information from neighbor nodes.
Combination function C: a function to combine aggregated information with the previous node representation.
Output: Updated model parameters theta (theta).
Steps:
Initialization:
Initialize model weights theta (theta) and noises.
For each labeled node v in Vl:
Set initial hidden representation h_v^theta (h_v superscript theta) to -alpha_l * theta(v).
This means the initial representation is the negative of alpha_l times the output of theta for node v.
For each unlabeled node u in Vu:
Set initial hidden representation h_u^theta (h_u superscript theta) to -alpha_u * theta(u).
This means the initial representation is the negative of alpha_u times the output of theta for node u.
Ascent Loop (M steps):
For t = 1 to M:
Update hidden representations for unlabeled nodes (aggregation and combination):
For each unlabeled node u in Vu:
Aggregate messages from neighbors:
msg_u^k (msg_u superscript k) is assigned A({(h_v^(k-1), h_u^(k-1), e_uv) | v belongs to N(u)})
This means message for node u at step k is the result of applying the aggregation function A to the set of tuples containing:
hidden representation of neighbor v at step k-1
hidden representation of node u at step k-1
the edge connecting node v and u
N(u) represents the set of neighbors of node u.
Combine aggregated message with previous hidden state:
h_u^k (h_u superscript k) is assigned C(h_u^(k-1), msg_u^k)
This updates the hidden representation of node u at step k by combining the previous representation at k-1 with the aggregated message using the combination function C.
Calculate loss and gradients:
L(h_v^k, y) (using backpropagation)
This computes the loss based on the hidden representation of labeled node v at step k and the true label y. Backpropagation is used to calculate gradients.
g_theta^t (g_theta superscript t) is assigned g_theta^(t-1) + grad_theta(L)
This updates the gradient of theta at step t by adding the gradient of the loss with respect to theta.
g_u^t (g_u superscript t) is assigned g_u^(t-1) + grad_u(L)
This updates the gradient of the hidden representation of unlabeled node u at step t by adding the gradient of the loss with respect to that representation.
Update parameters and noises:
theta is assigned theta + tau/M * g_theta^t
This updates the model parameters theta by adding the scaled gradient of theta at step t.
For each labeled node v in Vl:
delta_v^t (delta_v superscript t) is assigned delta_v^(t-1) + alpha_l * sign(grad_delta(L))
This updates the noise for labeled node v by adding alpha_l times the sign of the gradient of the loss with respect to the noise.
....
\end{lstlisting}
\end{listing}


\begin{listing}[!h]%
\caption{YONA - Prompt for SIngle LLM Call}%
\label{lst:listing2}%
\begin{lstlisting}
METHODOLOGY DESCRIPTION: 

{methodology}

STARTER CODE:

{starter_code}

RESNET CODE:

{resnet_code}

DATASET DESCRIPTION:

{data}

PSEUDOCODE:

{pseudocode}

STARTER CODE PERFORMANCE:

{starter_code_performance}


Edit the starter code for implementing the methodology using the information given above. 
\end{lstlisting}
\end{listing}

\begin{listing}[!h]%
\caption{YONA - Problem statement for ResearchCodeAgent}%
\label{lst:listing2}%
\begin{lstlisting}
You have been given files with the names methodology_description.txt, pseudocode.txt, model/resnet.py,dataset_description.txt, starter_code.py, starter_code_performance.txt.
Go through these files and understand their contents.
The file methodology_description.txt contains a detailed description of the YONA algorithm, which augments an image by splitting it into two equal pieces, applying data augmentation to one piece and replacing the other with noise, and then concatenating the pieces back together. This process leverages randomness to introduce diversity at both the patch and image levels.
The file pseudocode.txt contains pseudocode for the YONA method, which augments an image by randomly splitting it, applying data augmentation to one half, adding noise to the other half, and then concatenating them back together.
The file dataset_description.txt contains a description of the CIFAR-10 dataset, which consists of 60,000 color images in 10 classes, divided into training and test sets with 50,000 and 10,000 images respectively, and further split into batches for easier handling.
The file starter_code.py contains a starter code for the YONA algorithm which implements a PyTorch CIFAR training script with various options for dataset, optimizer, training epochs, and distributed training.You will need to first understand what this code does and also identify where you need to make changes to include the implementation of the FLAG algorithm. 
The file starter_code_performance.txt contains the performance of the starter code.
The working directory will contain an additional directory named models which contains code for Resnet.It contains code for implementing ResNet architectures, including ResNet18, ResNet34, ResNet50, ResNet101, and ResNet152, in PyTorch, based on the "Deep Residual Learning for Image Recognition" paper. Do not try to execute this file.
You can summarise the methodology_description.txt, dataset_description.txt, starter_code.py, starter_code_performance.txt and resnet.py in your research logs to keep track of what all you have to do.
Your task is to edit the code in starter_code.py to implement the methodology described in methodology_description.txt  and pseudocode.txt for the dataset described in dataset_description.txt.
Do not forget to execute the changes you made to check for performance and errors. Save the edited script in a new file named methodology_implementation.py.
Do not make any changes to the methodology. Do not change any arguments or hyperparameters from the starter code.
Make sure that you make appropriate edits to the code and execute the edited code and not just execute unedited code to submit the final answer.
Repair any errors that the code might have. Do not submit code with errors. Execute the code to verify that it does not have errors.
Use the Check Implementation action to check if the python script contains implementation of the methodology before submitting the final answer. If the implementation is incorrect then edit the code, execute the changes and verify that the implementation is correct before submitting the final answer.umber of times each line is executed and >>>>>> denotes that the line is not executed even once. 
\end{lstlisting}
\end{listing}

\begin{listing}[!h]%
\caption{YONA - CIFAR10 Dataset Description}%
\label{lst:listing2}%
\begin{lstlisting}
 The CIFAR-10 dataset consists of 60000 32x32 colour images in 10 classes, with 6000 images per class. There are 50000 training images and 10000 test images. Pixels store color information in red, green, and blue channels, with values from 0 (black) to 255 (bright).
The dataset is divided into five training batches and one test batch, each with 10000 images. The test batch contains exactly 1000 randomly-selected images from each class. The training batches contain the remaining images in random order, but some training batches may contain more images from one class than another. Between them, the training batches contain exactly 5000 images from each class.
\end{lstlisting}
\end{listing}

\begin{listing}[!h]%
\caption{YONA - Pseudocode}%
\label{lst:listing2}%
\begin{lstlisting}
Pseudocode for YONA Method

Define the image 'x' with dimensions (Channels, Height, Width)
Define a set of data augmentation functions 'A'

Procedure YONA_Augmentation(image, augmentation_functions):

    Step 1: Split the image 'x' into two equal segments either vertically or horizontally
    Get the dimensions of the image (Height, Width)
    Randomly choose to split the image vertically or horizontally
    Choose 'split_direction' randomly from ['vertical', 'horizontal']

    If split_direction is 'vertical':
        Split the image into two equal vertical segments
        Define 'mid_point' as Width divided by 2
        Segment1 = left half of the image up to 'mid_point'
        Segment2 = right half of the image from 'mid_point'
    Else:
        Split the image into two equal horizontal segments
        Define 'mid_point' as Height divided by 2
        Segment1 = top half of the image up to 'mid_point'
        Segment2 = bottom half of the image from 'mid_point'

    Step 2: Randomly select one segment to apply augmentation and add noise to the other
    If random value between 0 and 1 is less than or equal to 0.5:
        Augmented_Segment = Apply_Random_Augmentation(Segment1, augmentation_functions)
        Noised_Segment = Apply_Noise(Segment2)
        Combined_Image = Concatenate_Segments(Augmented_Segment, Noised_Segment, split_direction)
    Else:
        Augmented_Segment = Apply_Random_Augmentation(Segment2, augmentation_functions)
        Noised_Segment = Apply_Noise(Segment1)
        Combined_Image = Concatenate_Segments(Noised_Segment, Augmented_Segment, split_direction)

    Return Combined_Image

Procedure Apply_Random_Augmentation(segment, augmentation_functions):
    Step 3: Apply a randomly selected augmentation function to the segment
    Choose a random augmentation function from 'augmentation_functions'
    Apply the chosen augmentation function to 'segment'
    Return the augmented segment

Procedure Apply_Noise(segment):
    Step 4: Convert segment to Noise
    Generate random integer noise with the same range as the pixels of the image with the same dimensions as 'segment'
    Return the noised segment

Procedure Concatenate_Segments(segment1, segment2, split_direction):
    Step 5: Concatenate the two segments based on the split direction
    If split_direction is 'vertical':
        Concatenate segments along the width
        Return Concatenate 'segment1' and 'segment2' side by side
    Else:
        Concatenate segments along the height
        Return Concatenate 'segment1' and 'segment2' top to bottom

Example usage:
Define a list of augmentation functions as 'augmentation_functions'
Call YONA_Augmentation with the image and augmentation functions list 
\end{lstlisting}
\end{listing}

\begin{listing}[!h]%
\caption{YONA - Methodology Description}%
\label{lst:listing2}%
\begin{lstlisting}
YONA cuts one image into two equal pieces, either in the height or the width dimension. A specific data augmentation method is performed on one piece, and the pixels within the other piece are replaced with noise. Transformed pieces are then concatenated together to form one single augmented image.

Consider an image x of dimensions C*H*W and we let a() represent various data augmentation functions defined as a() : x' = a(x) and dimension of x' and x are exactly same. Here, x′ = a(x) denotes the augmented image. Unlike conventional image-level data augmentation methods that directly apply augmentations to the entire image as x′ = a(x), YONA initially bifurcates the image into two equal segments along either the height or width dimension, each selection occurring with equal probability.
possible examples for a() are Horizontal Flip (HFlip), Vertical Flip (VFlip), Color Jitter (Jitter), Random Erasing (Erasing).

[x1 x2] = cutH(x) if 0 <= p <= 0.5,
such that dimension of x1 and x2 = dim(x1) = dim(x2) = C*(H/2)*W
 or
[x1 x2] = cutW(x) if 0.5 <= p <= 1,
such that dimension of x1 and x2 = dim(x1) = dim(x2) = C*H*(W/2)

where p is sampled from (0, 1) uniformly (i.e., p blongs to U(0, 1)), x1 and x2 are cut pieces, cutH() and cutW () represent the cut operation in the height dimension and the width dimension, respectively.Then a() is applied to one randomly selected piece, and the pixels within the other piece are replaced with noise as:

xM1  = mask(x1) and 
xA2 = a(x2), if 0 <= q <= 0.5 
or
xA1 = a(x1) and 
xM2 = mask(x2), if 0.5 < q <= 1,

where q is sampled from from (0, 1) uniformly (i.e., q belongs to U(0, 1)), x1 and x2 are cut pieces, xM1 and XM2 are masked pieces, xA1 and xA2 are augmented pieces, mask() represents the masking with noise operation. Finally, we concatenate the transformed pieces back together as:

x' = concat[xM1, xA2], 
or
x' = concat[xA1, xM2],

where concat(,) represents the concatenation operation. Augmentation and masking operations are governed by randomness, encompassing random probabilities of being applied, random operations, and random magnitudes. These operations exhibit distinct behaviors, thereby instilling substantial diversity at both the patch and image levels.  
\end{lstlisting}
\end{listing} 

\begin{listing}[!h]%
\caption{OGSCL - Prompt for SIngle LLM Call}%
\label{lst:listing2}%
\begin{lstlisting}
METHODOLOGY DESCRIPTION: 

{methodology}

SUPPLEMENTARY METHODOLOGY:

{supplementary_methodology}

STARTER CODE:

{starter_code}

DATASET DESCRIPTION:

{data}

PSEUDOCODE:

{pseudocode}

STARTER CODE PERFORMANCE:

{starter_code_performance}

Edit the starter code for implementing the methodology using the information given above.
\end{lstlisting}
\end{listing}

 

\begin{listing}[!h]%
\caption{Problem statement for ResearchCodeAgent - OGSCL}%
\label{lst:listing2}%
\begin{lstlisting}
You have been given files with the names methodology_description.txt, pseudocode.txt, dataset_description.txt, starter_code.py, supplementary_methodology.txt, starter_code_performance.txt.
Go through these files and understand their contents.
The file methodology_description.txt contains a description of OGSCL, a method for fine-tuning CLIP for the fashion domain. OGSCL uses intelligent batching strategies that leverage a fashion ontology to create batches of images and captions with related attributes, improving attribute-based image understanding.
The file pseudocode.txt contains the OGSCL algorithm which processes a dataframe containing image attributes and captions to create a list of DataLoaders, each dedicated to a unique attribute type, along with the total number of batches across all DataLoaders.
The file dataset_description.txt contains a description of the DeepFashion dataset, which consists of fashion apparel images labeled with product categories and attributes like fabric, style, shape, texture, and part. This dataset is transformed into an image captioning dataset by creating captions that describe the attribute values of each product, resulting in multiple data points per image.
The file supplementary_methodology.txt contains a description of the methodology used in the starter code, which utilizes pre-trained CLIP on image-text pairs for supervised contrastive learning. It also explains how focal loss is incorporated to mitigate class imbalance issues encountered in datasets with rare attributes.
The file starter_code.py  contains starter code for the OGSCL algorithm, implementing a contrastive learning approach for image-caption alignment using the CLIP model and training and validation loops.You will need to first understand what this code does and also identify where you need to make changes to include the implementation of the OGSCL algorithm.
The file starter_code_performance.txt contains the performance of the starter code.
You can summarise the methodology_description.txt, dataset_description.txt, supplementary_methodology.txt, starter_code.py and starter_code_performance.txt in your research logs to keep track of what all you have to do.
Your task is to edit the code in starter_code.py to implement the methodology described in methodology_description.txt  and pseudocode.txt for the dataset described in dataset_description.txt.
Do not forget to execute the changes you made to check for performance and errors. Save the edited script in a new file named methodology_implementation.py.
Do not make any changes to the methodology. Do not change any arguments or hyperparameters from the starter code.
No other files apart from the ones mentioned here are available and are not necessary either.
Make sure that you make appropriate edits to the code and execute the edited code and not just execute unedited code to submit the final answer.
Repair any errors that the code might have. Do not submit code with errors. Execute the code to verify that it does not have errors.
Use the Check Implementation action to check if the python script contains implementation of the methodology before submitting the final answer. If the implementation is incorrect then edit the code, execute the changes and verify that the implementation is correct before submitting the final answer.  
\end{lstlisting}
\end{listing}

\begin{listing}[!h]%
\caption{OGSCL - Methodology Description}%
\label{lst:listing2}%
\begin{lstlisting}
OGSCL proposes an enhanced approach to fine-tuning CLIP for the fashion domain, leveraging multimodal representations and intelligent batching strategies to improve attribute-based image understanding. A fashion ontology consists of product category, attribute type and attribute label. Each fashion image belonging to a product category, such as shirt or blouse, is labelled an attribute value such as ‘pleated’ for an attribute type ‘fabric’. The captions for a fashion image are formed using the attribute type - value pairs. For example, the caption of an image of a blouse can be ‘The fabric of this Blouse is pleated’.  In the contrastive learning set-up, we define an image and a caption formed using an attribute value, the image is labelled with, as a positive pair.  Let AV11, AV12, AV13 are attribute values belonging to attribute type AT1, and thus are siblings of each other.  Let an image I1 is labelled with an attribute value AV11 belonging to attribute type AT1 forming a caption C11. Thus, (I1, C11)  forms a positive sample . Similarly let image I2 be labelled with AV13, forming (I2, C13) as a positive sample. For selecting in-batch negatives, we use the fashion ontology.  We explicitly make sure that in a batch only those image caption pairs are sampled, where the captions belong to the attribute values of the same attribute type. For example, as the part of intelligent batching,  (I1, C11)  is chosen as a sample then (I2,C13) is also chosen as a sample, because  AV1 and AV3 are siblings and belong to the same attribute type AT1.  Thus, in this batch (I1, C11)  and (I2,C13) are positive pairs and (I1, C13) and (I2, C11) form negative pairs. For example pleated,denim and knit are attribute values belonging to attribute type fabric and they would be considered siblings of one another.An image (Image1) labelled with attribute value pleated belonging to attribute type fabric has a caption - The fabric of this Blouse is pleated (Caption1). Thus this image and this caption form a positive sample and similarly and another image (Image2) with attribute value denim belonging to attribute type fabric has a caption - The fabric of this Cutoffs is denim (Caption2). This image and caption would also form a positive sample. In one of our intelligent batches we would take Image1 and Image2 as our images and Caption1 and Caption2 as our texts.This can be done because the attribute values of these images are siblings (siblings belong to the same attribute type) belonging to the attribute type fabric.The novelty of OGSCL is that it does not  change  the loss function at all but only implements intelligent batching where all the entries in the batch have the same attribute-type.
\end{lstlisting}
\end{listing}

\begin{listing}[!h]%
\caption{OGSCL - Dataset Description}%
\label{lst:listing2}%
\begin{lstlisting}
DeepFashion Description for OGSCL: The DeepFashion dataset comprises images of fashion apparel products along with their labels.Each sample product image in the dataset is labeled with a product category (PC) and associated attribute values (AV) for five different attribute types (AT): fabric, style, shape, texture, and part. Therefore, a typical label would be formatted as follows: [PC, ("fabric", AV1), ("style", AV2), ("shape", AV3), ("texture", AV4), ("part", AV5)]. It should be noted that a product image may not necessarily be labeled with all attribute types. The dataset includes a total of 983 attribute values. The dataset is then transformed into an image captioning dataset by applying the following procedure: for each (AT, AV) pair, a text caption is created in the format "The <AT> of this <PC> is <AV>." Consequently, in the new dataset, each image is repeated as many times as the number of (AT, AV) pairs it is labeled with. Thus, each data point in this transformed dataset will consist of an image and a corresponding caption that indicates the attribute value for a specific attribute type of the product.
The fashion Ontology in the dataset has the attribute types fabric, style, shape, texture and part, each having their own attribute values. The fashion Ontology is not explicitly required.
The dataset has the titles image, caption, category, attribute_type and attribute_value for its data points. They contain the path of the images, the caption of the given image which is verbalized category , attribute_type and attribute_value, the type of dress, attribute type and attribute value respectively. 
\end{lstlisting}
\end{listing}

\begin{listing}[!h]%
\caption{OGSCL - Pseudocode}%
\label{lst:listing2}%
\begin{lstlisting}
Algorithm: OGSCL

Input:
- dataframe: A data structure containing image attributes and associated captions
- image_path: Path to the directory containing images
- batch_size: Size of batches for DataLoader

Output:
- length: Total number of batches across all DataLoaders
- train_dataloader_list: List of DataLoaders for each unique attribute type

Procedure OGSCL(dataframe, image_path, batch_size)

1. Function process_dataframe(dataframe)
   - Initialize dataframe_list as an empty list
   - for each attribute_type in unique values of dataframe["attribute_type"] do
     - Filter dataframe to get filtered_dataframe where attribute_type matches
     - Append filtered_dataframe to dataframe_list
   - end for
   - return dataframe_list
   
2. Call process_dataframe(dataframe) to obtain train_dataframe_list

3. Initialize length to 0

4. Initialize train_dataloader_list as an empty list

5. for each train_dataframe in train_dataframe_list do
   - Create train_dataloader using:
     - Dataset from image_caption_dataset(train_dataframe, image_path)
     - batch_size set to batch_size
     - Shuffle enabled
   - Add the number of batches in train_dataloader to length
   - Append train_dataloader to train_dataloader_list
   
6. end for

7. return length and train_dataloader_list
\end{lstlisting}
\end{listing}

\begin{listing}[!h]%
\caption{OGSCL - Supplementary methodology}%
\label{lst:listing2}%
\begin{lstlisting}
This methodology is already implemented in the starter_code.py and is only present for your context.
CLIP is pre-trained on (image, text) pairs by maximizing the cosine similarity between representations of B (image, text) pairs and minimizing the cosine similarity for B^2 − B invalid pairs in a batch of size B.
We obtain the multimodal image representation by first passing the image through CLIP's image encoder, followed by a multimodal projection layer. Similarly, we use the caption provided in the dataset, which verbalizes the attribute value, obtaining the corresponding multimodal representation via CLIP’s text encoder and a multimodal projection layer.
During the fine-tuning of CLIP, we maximize the cosine similarity between the representations of positive (image, attribute value) pairs and minimize the cosine similarity for negative pairs. 
\end{lstlisting}
\end{listing}

\end{document}